\documentclass[useAMS,fleqn,usenatbib]{mnras}
\usepackage{natbib}
\usepackage{mathtools}
\usepackage[T1]{fontenc}
\usepackage{ae,aecompl}
\usepackage{graphicx}
\usepackage{amsmath}
\usepackage{amssymb}
\usepackage{txfonts}

\title[Ultra-compact dwarf galaxies]{A consistency-test for determining
  whether ultra-compact dwarf galaxies could be the remnant nuclei of threshed
  galaxies}

\author[Graham]{Alister W.\ Graham\thanks{E-mail: AGraham@swin.edu.au}
\\
Centre for Astrophysics and Supercomputing, Swinburne University of
Technology, Hawthorn, VIC 3122, Australia. 
}

\date{Accepted XXX. Received YYY; in original form ZZZ}
\pubyear{2019}

\begin{document}
\label{firstpage}
\pagerange{\pageref{firstpage}--\pageref{lastpage}}
\maketitle

\begin{abstract}

It has been suggested that ultra-compact dwarf (UCD) galaxies are the ``threshed''
remains of larger galaxies.  Simulations have revealed that extensive
tidal-stripping may pare a galaxy back to its tightly-bound, compact nuclear
star cluster.  It has therefore been proposed that the two-component nature of
UCD galaxies may reflect the original nuclear star cluster surrounded by the paltry
remnants of its host galaxy.  A simple quantitative test of this theory is
devised and applied here.  If the mass of the central black hole in UCD galaxies,
relative to the mass of the UCD galaxies' inner stellar component, i.e.\ the suspected nuclear
star cluster, matches with the (black hole
mass)-(nuclear star cluster mass) relation observed in other galaxies, then it would
provide quantitative support for the stripped galaxy scenario.  Such 
consistency is found for four of the five UCD galaxies reported to have a massive
black hole.  This (black hole mass)-(nuclear star cluster mass) 
relation is then used to predict the central black hole
mass in two additional UCD galaxies, and to reveal that NGC~205 and possibly NGC~404
(which only has an upper limit to its black hole mass) 
also follow this scaling relation. 

\end{abstract}


\section{Introduction}

Several origins for ultra-compact dwarf (UCD) galaxies
\citep{1995ApJ...441..120H, 1999A&AS..134...75H, 2000PASA...17..227D,
  2006A&A...447..877G} have been proposed.  The apparent lack of dark matter
in UCD galaxies \citep[e.g.][]{2007A&A...463..119H, 2011MNRAS.412.1627C,
  2011MNRAS.414L..70F} ruled out the notion of them being compact galaxies
formed in small, compact dark matter halos.  This left alternatives such as:
large globular clusters possibly built over two or multiple epochs
\citep[e.g.][]{2000IAUJD...5E...4N, 2004ApJ...605L.125B, 2014MNRAS.444.3670P};
globular cluster mergers \citep[e.g.][]{1997ApJ...487L.187N}; compact young
massive clusters formed during past galaxy interactions
\citep[e.g.][]{1998MNRAS.300..200K, 2004A&A...416..467M, 2017ApJ...843...91L,
  2017ApJ...844..108M}; direct formation of star clusters from supergiant
molecular clouds \citep{2018MNRAS.478.3564G}; failed galaxies surrounding the
remnant black hole of a Population III star or a primordial black hole
\citep{2017JCAP...04..036D}; or perhaps the remnant nuclei of
tidally-stripped, low-mass galaxies \citep{1988IAUS..126..603Z,
  1993ASPC...48..608F, 2001ApJ...552L.105B, 2003MNRAS.344..399B,
  2003Natur.423..519D, 2009MNRAS.396.1075G, 2015MNRAS.451.3615N,
  2015ApJ...802...30Z}.  The latter scenario is of interest because it can be
tested.

\citet{2001ApJ...557L..39B} 
show how an early-type disc galaxy can have much of its stellar
disc, and some of its bulge, stripped away by a massive neighbour to form a
``compact elliptical'' dwarf galaxy (e.g., M32 and NGC~4486B), of which at
least one is now known to have an active galactic nucleus
\citep{2016ApJ...820L..19P}. 
The stripping explains why these rare\footnote{While grossly over-represented
  in some published scaling diagrams, at a given luminosity, just 0.5 percent
  of dwarf galaxies are ``compact elliptical'' galaxies (Chilingarian, private
  communication).} galaxies are 
overly metal rich for their luminosity 
\citep[e.g.][]{2009Sci...326.1379C, 2009MNRAS.397.1816P}, 
underluminous for their colour 
\citep[e.g.][, their Figure~11, and references therein]{2019MNRAS.484..794G}, 
and should be excluded when establishing the 
$M_{\rm bh}$--$M_{\rm galaxy}$ and 
$M_{\rm bh}$--$M_{\rm bulge}$ scaling relations. 
Further stripping might reduce a galaxy 
to its more tightly-bound, nuclear star cluster 
\citep{2008MNRAS.385L..83C, 2012ApJ...755L..13K, 2013MNRAS.433.1997P}. 
\citet{2003MNRAS.346L..11B} and \citet{2004ApJ...616L.107I} 
investigated
how dwarf elliptical and dwarf disc galaxies might be pared down by strong 
external gravitational tides, leaving an object similar to $\omega$ Centauri
and the UCD galaxies.

The commonly-observed, two-component 
nature of a UCD galaxies' structure might reflect the original nuclear star 
cluster surrounded by the some of the remnants of its former host galaxy 
\cite[e.g.][]{2007AJ....133.1722E, 2015ApJ...812L..10J, 2016A&A...586A.102V,
  2016MNRAS.459.4450W}. 
Such objects would still house the original galaxy's supermassive black hole (SMBH), or
perhaps intermediate-mass black hole (IMBH: $10^2 < M_{\rm bh}/M_{\odot} <
10^5$).  \citet{2009MNRAS.396.1075G} 
have shown that the nuclear star clusters in (mainly) 
late-type dwarf galaxies resemble massive globular clusters, which they note might not
be globular clusters but nuclei largely stripped of their former galaxy
\citep{2008ApJ...672L.111B}. 


There has been an abundance of claims, usually followed by counter-claims,
for the existence of IMBHs in large globular clusters.  For example, see the studies of:
G1, the most massive globular cluster around M31 \citep{2005ApJ...634.1093G,
  2012ApJ...755L...1M}; 
$\omega$ Centauri, the most massive globular
cluster around the Milky Way \citep{2008ApJ...676.1008N, 2010ApJ...710.1032A,
  2010ApJ...710.1063V, 2013ApJ...773L..31H, 2017MNRAS.468.4429Z,
  2019MNRAS.482.4713Z}; 
plus other globular clusters around the Milky Way such as M15
\citep{2000AJ....119.1268G, 2002AJ....124.3270G, 2003ApJ...582L..21B,
  2012A&A...542A..44K, 2014MNRAS.438..487D, 2014A&A...565A..43K}; 
M54 \citep{2009ApJ...699L.169I}; 
47~Tucanae \citep{2017Natur.542..203K, 2018MNRAS.481..627A,
  2019ApJ...875....1M}; 
and NGC~6624 \citep{2017MNRAS.468.2114P, 2018MNRAS.473.4832G,
  2019MNRAS.488.5340B}.  
To date, there remains no conclusive evidence for IMBHs in globular clusters
\citep{2007MNRAS.379...93H, 2010ApJ...720L.179V, 2016MmSAI..87..563L,
  2016MmSAI..87..559M, 2016AJ....152...22W, 2018Msngr.172...18F,
  2018ApJ...862...16T}, although see \citep{2016IAUS..312..181L}. 
This current status may simply reflect the observational difficulty in detecting low mass
black holes in a relatively gas poor environment. 
While few IMBHs are currently known 
\citep[e.g.][]{2009Natur.460...73F, 2015ApJ...809L..14B, 2016ApJ...818..172G,
  2018ApJ...863....1C}, 
many have recently been predicted at the
centres of low-mass galaxies in the Virgo cluster 
\citep{2019MNRAS.484..794G, 2019MNRAS.484..814G}, with follow-up {\it Chandra} X-ray data
in tow (PI: R.~Soria. Proposal ID: 18620568), and beyond 
\citep{2019arXiv191202860M}.  Given that UCD galaxies may represent the 
nuclei of low-mass galaxies, UCD galaxies may thus be good targets for IMBH research
\citep[e.g.][]{2014AJ....148..136M, 2015MNRAS.454.3722S, 2018ApJ...863....1C,
  2019MNRAS.484..814G, 2019ApJ...884...54M}. 

Nuclear star clusters 
\citep[e.g.][]{2002AJ....123.1389B, 2004AJ....127..105B, 2012MNRAS.424.2130L} 
differ from globular star clusters in that they reside
at the bottom of the gravitational potential well of their host galaxy, which
can help to retain stellar winds and funnel down
all manner of material that has lost its orbital angular momentum either
through shocks in the case of gas 
\citep[e.g.][]{1999AJ....118.2646M, 2004ApJ...605L..13M}, 
or dynamical friction in the case of denser bodies
\citep[e.g.][]{1943ApJ....97..255C, 1975ApJ...196..407T, 2011MNRAS.416.1181I,
  2014ApJ...785...51A}.  For over a decade, 
many nuclear star clusters with supermassive black
holes have been known to coexist 
\citep[e.g.][and references therein]{2008AJ....135..747G, 2008ApJ...678..116S,
  2008ApJ...682..104S, 2009MNRAS.397.2148G}.  
The Milky Way and M32 were among the first galaxies recognised to house both 
\citep{1969Natur.223..690L, 1972AJ.....77..292S, 1984ApJ...283L..27T}, and 
other examples are reported in  
\citet{2012AdAst2012E..15N, 2016MNRAS.457.2122G}. 
If UCD galaxies are indeed remnant nuclear star clusters, after their host
galaxy has largely been removed, then some may harbour a massive black hole. 


\begin{figure}
 \includegraphics[angle=-90, trim=2.2cm 3cm 1cm 6.8cm,
   width=\columnwidth]{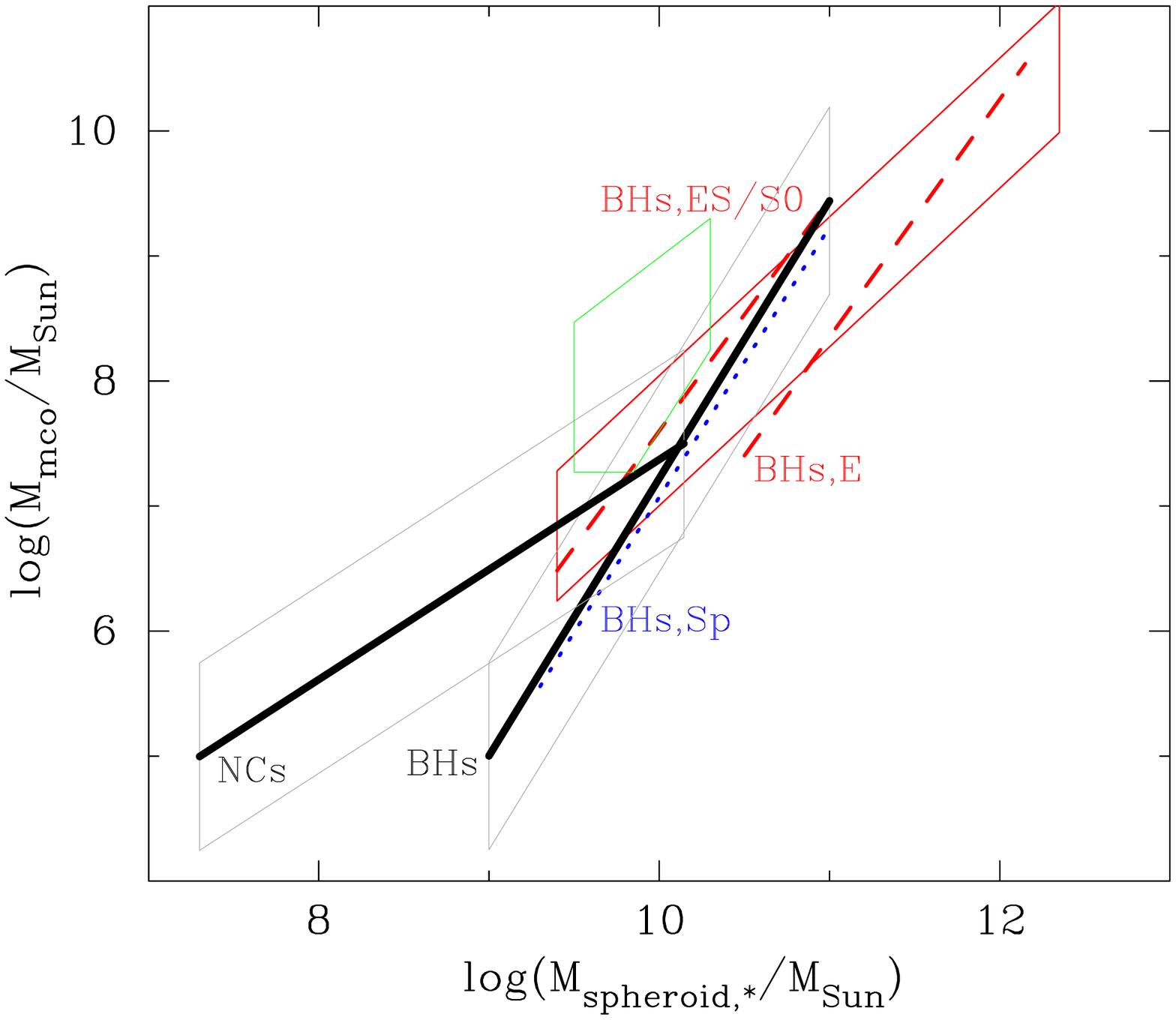}
    \caption{$M_{\rm cmo}$--$M_{\rm spheroid,*}$ diagram, where the central
      massive object (cmo) may be a nuclear star cluster (NC: 
      black line, equation~\ref{Eq_xyz}), a larger, flatter nuclear disc (ND: region
      denoted by the green pentagon), or a supermassive black hole (BH: black
      line,  
      equation~\ref{eq_Sersic}, for galaxies with large- or intermediate-scale discs). 
The three (morphological 
      type)-specific black hole relations shown by the blue dotted line
      and the two red dashed lines have been taken from 
\citet{2019ApJ...876..155S}. 
  They represent spiral (Sp) galaxies, ellicular \citep[ES, see][]{2019PASA...36...35G}
      and lenticular (S0) galaxies, 
      and discless elliptical (E) galaxies.  When combined, the E, ES and S0
      galaxies occupy the region traced by the red parallelogram. 
    }
\label{Fig1}
\end{figure}

Obviously, given the array of formation scenarios noted above, 
the presence of a massive black hole in a UCD galaxy does not imply that
they are the nuclear star clusters of stripped galaxies.  However,
if the (black hole)-to-(inner star cluster) mass ratio in UCD galaxies matches that
observed in nuclear star clusters, it does then become suggestive of this
scenario.  
The various scaling relations connecting supermassive black holes with the
different physical properties of their host bulges were reviewed in
\citet{2016ASSL..418..263G}, with references to many key, but often over-looked, papers.
The relation between the logarithm of the black hole mass and the logarithm of
the host bulge mass now extends down to black hole masses of $10^5~M_{\odot}$,
and bulge masses of $10^9~M_{\odot}$ \citep{2015ApJ...798...54G}. 
The super-quadratic $M_{\rm bh}$--$M_{\rm bulge}$ 
relation observed at low masses has since been confirmed for spiral
galaxies \citep[which are all S\'ersic galaxies, as opposed to core-S\'ersic
galaxies with partially depleted stellar cores:][]{2003AJ....125.2951G} 
using the latest samples with 
directly measured black hole masses and careful multicomponent decompositions
of the galaxy light \citep{2016ApJ...817...21S, 2019ApJ...873...85D}. 
This revised view, captured in \citet{2019ApJ...876..155S}, 
see also \citep{2000MNRAS.317..488S, 2001ApJ...553..677L,
  2012ApJ...746..113G}, 
departs dramatically from the early picture of a single near-linear $M_{\rm
  bh}$--$M_{\rm bulge}$ 
relation \citep{1989IAUS..134..217D, 1995ARA&A..33..581K, 1998AJ....115.2285M,
  2013ARA&A..51..511K}, see Figure~\ref{Fig1}. 


Low-mass (non dwarf
spheroidal)\footnote{Dwarf spheroidal galaxies are fainter than dwarf elliptical
  galaxies, and have long been known not to be particularly 
nucleated \citep{1959PDDO....2..147V, 1973Ap......9...33M}.} 
galaxies tend to have nuclear star 
clusters rather than depleted cores 
\citep[e.g.][]{1985AJ.....90.1759S, 2003AJ....125.2936G, 2003ApJ...582L..79B,
  2007ApJ...665.1084B, 2006ApJS..165...57C, 2006ApJS..164..334F}.  
Not surprisingly, mass scaling relations for nuclear star clusters and their
host bulge also exist, although they have not received the same attention as the 
black hole mass scaling relations. 
Combining the low-mass 
$M_{\rm bh}$--$M_{\rm bulge}$ and $M_{\rm nc}$--$M_{\rm bulge}$ 
relations to cancel out the 
bulge mass yields a relation between the black hole and nuclear
cluster mass \citep{2016IAUS..312..269G}.   
One can then test if UCD galaxies
do, or do not, follow this relation. If they do, it will provide quantitative 
support for the threshing scenario. 

Section~\ref{Sec_meth} describes this method in more detail, and the use of
the $M_{\rm bh}$--$\sigma$ and $M_{\rm nc}$--$\sigma$ relations to construct
an additional $M_{\rm bh}$--$M_{\rm nc}$ relation that can be, and is, used.
Section~\ref{Sec_Results} applies this to all five UCD galaxies with a
published black hole mass and available inner-stellar-component mass.  The
black hole mass is also predicted for an additional two UCD galaxies with
available inner-component stellar masses.  In Section~\ref{Sec_nc}, two
additional nuclear star clusters --- in one dwarf galaxy recently reported
to have a black hole, and in another with an upper limit on the black hole mass --- are checked
to see if they too conform to the $M_{\rm bh}$--$M_{\rm nc}$ relation. 
A brief discussion is provided in Section~\ref{Sec_discuss}, while 
noting some of the many implications arising from UCD galaxies with massive
black holes.

\section{Methodology}\label{Sec_meth}

The test herein is straightforward, bar one twist that we shall get to,
and which surprisingly has not been performed for UCD galaxies. 
It inputs the published black mass of the UCD galaxy 
into the $M_{\rm bh}$--$M_{\rm nc}$ scaling relations (see below) 
to calculate the expected host nuclear star cluster mass. 
This expected mass is then compared with the observed 
mass of the UCD galaxies' inner-component, i.e., the suspected nuclear star cluster.
If UCD galaxies are not the threshed remains of a galaxy, 
but formed via a different formation channel (for example, failed
galaxies around SMBH seeds), 
then the predicted nuclear cluster mass and the observed
inner-cluster mass may not agree.  Put another way, if the masses 
are found to disagree, it would disfavour the threshing origin of UCD galaxies, while
agreement would offer support for such a scenario. 

Following \citet{2016IAUS..312..269G}, 
the $M_{\rm nc}$--$M_{\rm spheroid}$ relation 
can be united with the 
$M_{\rm bh}$--$M_{\rm spheroid}$ relation (for galaxies without depleted
cores) to eliminate the quantity 
$M_{\rm spheroid}$ and produce an $M_{\rm nc}$--$M_{\rm bh}$ scaling relation. 
The $M_{\rm nc}$--$\sigma$ and
$M_{\rm bh}$--$\sigma$ relations can also be combined to eliminate the quantity
$\sigma$ and yield an additional $M_{\rm nc}$--$M_{\rm bh}$ scaling relation.
This is done here.

First, combining the relation 
\begin{equation}
\log(M_{\rm nc}/M_{\odot}) = (0.88\pm0.19)\log(M_{\rm sph}/[10^{9.6}
  M_{\odot}]) + (7.02\pm0.10) 
\label{Eq_xyz}
\end{equation}
from \cite{2013ApJ...763...76S} 
together with the relation 
\begin{equation}\label{eq_Sersic}
\log(M_{\rm bh}/M_{\odot}) = (2.22\pm0.58)\log(M_{\rm 
  sph}/[2\times10^{10} M_{\odot}]) +(7.89\pm0.18)
\end{equation}
from \citet{2013ApJ...768...76S} 
for the S\'ersic, i.e.\ not core-S\'ersic, galaxy sample 
--- which is consistent with the 
$M_{\rm bh}$--$M_{\rm sph}$ relation for spiral galaxies from
\citet{2019ApJ...873...85D} and \citet{2019ApJ...876..155S}, 
and which has been shown to extend to black hole
masses of $10^5\,M_{\odot}$ \citep{2015ApJ...798...54G} --- 
gives our first $M_{\rm bh}$--$M_{\rm nc}$ relation: 
\begin{eqnarray}
&& \log(M_{\rm nc}/M_{\odot}) = (0.40\pm0.13)
\log(M_{\rm bh}/[10^{7.89}\,M_{\odot}]) + (7.64\pm0.18)  \\
\label{Eq_mass}
\end{eqnarray}
One advantage with the above combination is that any systematic errors in the
stellar mass of the spheroid should cancel out. 

Building on \cite{2012MNRAS.422.1586G}, \citet{2013ApJ...763...76S}, 
\citep[see also][]{2017MNRAS.472.4013C} reported that 
\begin{equation}
\log(M_{\rm nc}/M_{\odot}) = (2.11\pm0.31)\log(\sigma/54) + (6.63\pm0.09). 
\end{equation}
Combining this with the relation 
\begin{equation}
\log(M_{\rm bh}/M_{\odot}) = (5.53\pm0.34)\log(\sigma/200) + (8.22\pm0.05) 
\end{equation} 
for non-barred galaxies from \citet[][, their Table~3]{2013ApJ...764..151G} 
--- which is consistent with the
symmetric $M_{\rm bh}$--$\sigma$ relations in 
\citet{2019ApJ...887...10S}\footnote{\citet{2019ApJ...887...10S}
observe consistent $M_{\rm
    bh}$--$\sigma$ relations for their larger sample of barred and
  non-barred galaxies.} --- 
gives:  
\begin{eqnarray} 
&& \log(M_{\rm nc}/M_{\odot}) = (0.38\pm0.06)
\log(M_{\rm bh}/[10^{8.22}\,M_{\odot}]) + (7.83\pm0.20) \nonumber  \\
&& = (0.38\pm0.06)
\log(M_{\rm bh}/[10^{7.89}\,M_{\odot}]) + (7.70\pm0.20)  \\
\label{Eq_sigma}
\end{eqnarray}

Inverting this equation, one obtains $M_{\rm bh} \propto M_{\rm
  nc}^{2.6\pm0.4}$, consistent with the exponent 2.7 reported in 
\citet{2016IAUS..312..269G}. 
These relations involving nuclear star cluster mass are observed over
the range $10^5 \lesssim M_{\rm nc}/M_{\odot} \lesssim 5\times10^7$ and with
half light radii less than $\approx$20 pc and generally around 4~pc.  More
massive nuclear star clusters likely blend into the population of nuclear
discs, which depart from these scaling relations \citep[see][their Figures~1
  and 2]{2013ApJ...763...76S}. 

In the $\log M_{\rm bh}$--$\log M_{\rm nc}$ diagram, 
a slope of 2.7 will span 7.3 orders of magnitude in black hole mass across the
above mentioned 2.7 orders of magnitude in nuclear cluster mass. 
It also corresponds to an $M_{\rm bh}/M_{\rm nc}$ mass ratio
that increases by 4.6 orders of magnitude from 
 $M_{\rm nc}/M_{\odot} = 10^5$ to $5\times10^7\,M_{\odot}$.  At this
higher-mass juncture in the $M_{\rm
  bh}$--$M_{\rm nc}$ diagram, two different evolutionary scenarios can be
envisioned, leading to a bifurcation.   A dry merger event may bring in a
second massive black hole (and star cluster) forming a massive BH pair which
scours away the 
nuclear star cluster(s) along with the core of the host galaxy
\citep{2010ApJ...714L.313B, 1980Natur.287..307B}. 
Alternatively, a wet merger or accretion event may build a larger, flatter, 
nuclear disc (creating the upturn seen in Figures~1 and 2 in
\citet{2013ApJ...763...76S}; see also \citet{2017ApJ...849...55S} and
\citet[][, their Figure~4]{2019ApJ...878...18S}).\footnote{This 
is not meant to imply a discontinuity between star
  clusters and nuclear discs.}  These two processes may also account
for some of the scatter in the $M_{\rm bh}$--$M_{\rm nc}$ relation observed
from $5 < \log M_{\rm nc}/M_{\odot} < 7.7$. 
 
The above non-linear $M_{\rm bh}$--$M_{\rm nc}$ scaling relations imply a
range of (black hole)-to-(nuclear star cluster) mass ratios.  This is evident
by comparing the Milky Way and M32, with their similar mass ratio of 1-to-10,
with galaxies like NGC~1023 that have a mass ratio of 10-to-1, or NGC~3115
with a ratio approaching 100-to-1
\citep[e.g.][]{2009MNRAS.397.2148G}. Furthermore, \citet{2016MNRAS.457.2122G}
report a range of ratios from $10^{-4}$ to $10^4$ when including nuclear
discs.  If most massive UCD galaxies host high-mass fraction BHs
\citep{2017ApJ...839...72A}, then UCD galaxies would appear to be inconsistent
with the notion that they are the tidally-stripped, remnant nuclear star
clusters of galaxies which have a broad range of (black hole)-to-(nuclear star
cluster) mass ratios.

\section{Results}\label{Sec_Results}

\subsection{M60-UCD1}

The measurement of the black hole mass in M60-UCD1 comes from 
\citet{2014Natur.513..398S}, and is 
$(2.1\pm0.4)\times10^7 M_{\odot}$ (1$\sigma$ uncertainty).  This is reported there to be 15\% of the
total mass, which is thus $1.4\times10^8 M_{\odot}$.

The $g$-band and $z$-band luminosity ratio of the inner-component
relative to the total luminosity is 0.58 \citep[][, their table~1]{2013ApJ...775L...6S}. 
Assuming this luminosity comes from stellar light, 
a 0.58 fraction of 85\% of the total mass yields a stellar mass for the 
inner-component equal to $6.9\times10^7 M_{\odot}$. 
This calculation assumes equal stellar mass-to-light ratios for the inner- and
outer-component of M60-UCD1.  This gives a mass ratio of the black hole
relative to the inner component of $2.2 / 6.9$ or about one-third. 

In passing, it is noted that the \citet{1968adga.book.....S} 
index \citep[see][for a review of the S\'ersic model]{2005PASA...22..118G} 
of the inner component is 3.3. 
This compares favourably with the S\'ersic index of the nuclear star cluster
in the Milky Way ($n=3$) and in M32 ($n=2.3$) (\citet{2009MNRAS.397.2148G};
see also \citet{2019ApJ...872..104N} who report $n=2.7\pm0.3$ for M32). 

Plugging in $M_{\rm bh} = (2.2\pm0.4)\times10^7 M_{\odot}$ 
\citep{2014Natur.513..398S} into equation~\ref{Eq_mass}, 
and assuming an intrinsic scatter of 0.5 dex in the $\log M_{\rm nc}$
direction, gives 
$\log(M_{\rm nc}/M_{\odot}) = 7.42\pm0.53$, 
or $M_{\rm nc} = (2.6^{+6.3}_{-1.8})\times10^7 M_{\odot}$. 
This predicted nuclear cluster mass is just 2.65 times smaller 
than the stellar mass of the inner component in M60-UCD1 
($=6.9\times10^7 M_{\odot}$), i.e.\ the suspected nuclear star cluster.  

Plugging $M_{\rm bh} = (2.2\pm0.4)\times10^7 M_{\odot}$ into equation~\ref{Eq_sigma},
and again assuming an intrinsic scatter of 0.5 dex in the $\log M_{\rm nc}$
direction, gives the consistent result 
$\log(M_{\rm nc}/M_{\odot}) = 7.49\pm0.54$,
or $M_{\rm nc} = (3.1^{+7.6}_{-2.2})\times10^7 M_{\odot}$.

This simple test, which was not previously performed for M60-UCD1, provides 
support for the stripped-galaxy scenario, in which a threshing process may
leave behind the dense central seed (nuclear cluster plus black hole) of the pared galaxy.

\subsection{Virgo cluster: VUCD3 and M59cO}

Following \citet{2014Natur.513..398S}, 
\citet{2017ApJ...839...72A} 
claimed the detection of massive black holes in two UCD
galaxies (VUCD3 and M59cO) residing in the Virgo galaxy cluster.  
VUCD3 is particularly interesting given that \citet{2015ApJ...812L...2L} had detected
a possible tidal stream associated with this target. 
While \citet{2017ApJ...839...72A}
noted that high levels of orbital anisotropy in these two 
UCD galaxies could nullify their suspected signature 
of a black hole, they reported 
$M_{\rm bh}=4.4^{+2.5}_{-3.0}\times10^6~M_{\odot}$ in VUCD3 and
$M_{\rm bh}=5.8^{+2.5}_{-2.8}\times10^6~M_{\odot}$ in M59cO
(3$\sigma$ uncertainties). 

As done above for M60-UCD1, we can employ (the more accurate) 
equation~\ref{Eq_sigma} to determine what the expected mass
should be for the inner-component of VUCD3 and M59cO if they are the remnant nuclear star clusters of
galaxies.  Doing so, one obtains stellar mass expectations of 
$\log(M_{\rm nc}/M_{\odot}) = 7.23\pm0.56$ and $7.27\pm0.55$, 
or $M_{\rm nc} = (1.7^{+4.4}_{-1.2})\times10^7 M_{\odot}$ and 
$(1.9^{+4.7}_{-1.4})\times10^7 M_{\odot}$, 
respectively.

These masses, based upon the black hole masses, can be compared
with the stellar masses reported by \citet[][, their Section~3]{2017ApJ...839...72A}
for the inner-component of their two-component decompositions of each UCD
galaxies' image. For VUCD3, they reported $M_{\rm
  *,inner}=(1.1\pm0.3)\times10^7~M_{\odot}$, while for M59cO they reported
$M_{\rm *,inner}=(1.4^{+0.2}_{-0.2})\times10^7~M_{\odot}$.  This remarkable
agreement, in both UCD galaxies, further supports the notion that
these UCD galaxies could be the nuclei and remnants of threshed galaxies.  
Once again, this test had never been performed for these two UCD galaxies. 

The $M_{\rm 
  bh}$-to-$M_{\rm *,inner-cpt}$ mass ratios are $\sim$40 percent in both of these
UCD galaxies. The smaller $M_{\rm bh}$-to-$M_{\rm *,UCD}$ ratio is of course somewhat
irrelevant, depending simply on how much of the host galaxy remains around the
original nuclear star cluster. 
Using the mass of the suspected former nuclear star clusters, 
equation~\ref{Eq_xyz} suggests a former spheroid/bulge mass of 4 and
5$\times10^{9}~M_{\odot}$ 
with an uncertainty of a factor of $\sim$4 which is 
dominated by the intrinsic scatter in the $M_{\rm nc}$--$M_{\rm *,sph}$ relation. 
While these estimates are several times more massive than the bulge mass estimates of 1.2
and 1.7$\times10^{9}~M_{\odot}$ in \citet{2017ApJ...839...72A} 
 --- arising from their use of a near-linear $M_{\rm bh}$--$M_{\rm *,sph}$
scaling relation which does not reach these low black hole masses until lower bulge
masses --- the uncertainty in these bulge mass estimates results in
consistency.


\subsection{M59-UCD3}

A fourth Virgo cluster UCD galaxy reported to have a massive black hole is M59-UCD3
\citet{2018ApJ...858..102A}. 
Although their triaxial Schwarzschild model has a minimum
$\chi^2$ value at a black hole mass of zero, from their axisymmetric
Schwarzschild model, and Jeans anisotropic model (JAM), they determine a black
hole mass of $4.2^{+2.1}_{-1.7}\times10^6~M_{\odot}$ (1$\sigma$
uncertainties).  From equation~\ref{Eq_sigma}, this would suggest a host star
cluster of stellar mass given by
$\log {\rm M}_{\rm nc} = 7.22\pm0.55$, or 
${\rm M}_{\rm nc}=1.7^{4.2}_{-1.2} \times 10^7\,{\rm M}_{\odot}$.

\citet{2018ApJ...858..102A} decomposed the image of M59-UCD3 into three components.
The inner ($R_{\rm e,inner}=16.8$ pc), middle ($R_{\rm e,middle}=40.0$ pc),
and outer ($R_{\rm e,outer}=99.2$ pc) components were reported to have a
stellar mass of
$(1.73\pm0.28) \times 10^8\,{\rm M}_{\odot}$, 
$(1.06\pm0.15) \times 10^8\,{\rm M}_{\odot}$, 
and $(0.12\pm0.1) \times 10^8\,{\rm M}_{\odot}$, 
respectively.  The inner-component has a mass which is an order or magnitude
greater than expected from the above calculation.  Oddly, the inner-component also has an axis ratio of
0.74 while the outer components are round rather than stretched out. The
colour of the inner- and middle components are rather similar, and
significant rotation (24-30 km s$^{-1}$) is observed across the inner $\sim$40
pc.  Conceivably, M59-UCD3 may represent the remnant nuclear disc of a
threshed galaxy.  \cite{2007ApJ...665.1084B} have reported on the commonality of
nuclear discs (tens of parsecs in size) in early-type disc galaxies.  Given
that the scaling relations in Section~\ref{Sec_meth} were established for
nuclear star clusters (with masses less than $5\times10^7\,{\rm M}_{\odot}$),
  it may therefore be inappropriate to apply the consistency test in this
  instance, as it is not designed to predict the masses of nuclear discs, or
  it may represent a failure of the consistency test.

\subsection{Fornax-UCD3} 

Fornax-UCD3 \citep{2011MNRAS.414L..70F} was observed with a {\it Hubble
  Space Telescope} / Advanced Camera for Surveys {\it F606W} image and 
modelled as the sum of two S\'ersic components by
\citet{2008AJ....136..461E}. 
However, the magnitude of the inner S\'ersic component was not reported, only the
combined/total UCD galaxy magnitude was given, along with $R_{\rm e,total}=86.5\pm6.2$~pc ($0.\arcsec94$).  
It was also noted there that a possible spiral galaxy in the background of
Fornax-UCD3 complicated the analysis. 

In this UCD galaxy, \citet{2018MNRAS.477.4856A} 
have reported a BH mass similar to that
in M59-UCD3 and equal to $3.3^{+1.4}_{-1.2}\times 10^6\,M_{\odot}$ (1-sigma
uncertainty), from which one would expect, using equation~\ref{Eq_sigma}, a
nuclear star cluster stellar mass given by $\log {\rm M}_{\rm nc} =
7.18\pm0.55$, or ${\rm M}_{\rm nc} = 1.5^{+3.9}_{-1.1} \times 10^7\,{\rm
  M}_{\odot}$. 

\citet{2018MNRAS.477.4856A} fit a two-component model to Fornax-UCD3, reporting
an $F606W$ magnitude of 20.17 (AB mag) for the inner component, to which they
apply a stellar mass-to-light ratio of 3.35. This gives a stellar mass of
$9.7\times10^6\,{\rm M}_{\odot}$, for their adopted distance of 20.9 Mpc, in
good agreement with the predicted mass.

\subsection{Centaurus~A: UCD~320 and UCD~330}

\citet{2018ApJ...858...20V}, see also \citet{2007A&A...469..147R}, 
have observed two low-mass UCD galaxies 
(UCD~320 and UCD~330: $M_{\rm UCD} < 10^7 M_{\odot}$), first discovered by
\citet{1992AJ....104..613H}, around the galaxy Centaurus~A (NGC~5128).
They did not, however, detect the presence of a central black hole in either
of these galaxies, and thus there are, to date, just five UCD galaxies with a reported
black hole mass measurement. 

Nontheless, for UCD~330, they reported 
that the absolute magnitude is $M_V = -11.03$~mag, and that $M_*/L_V = 3.30$,
giving a total stellar mass of $7.15\times10^6~M_{\odot}$ when using
$M_{\odot,V}=+4.81$~mag. 
They additionally fit a two-component structure to the distribution of light
in UCD~330, reporting that 62.4 per cent of the light resides in the inner component,
which therefore has a stellar mass equal to $4.46\times10^6~M_{\odot}$. 
From equation~\ref{Eq_new-sigma}, which represents the inversion of
equation~\ref{Eq_sigma}.
the expected black hole mass based on this
speculated nuclear star cluster mass is $1.3\times10^5~M_{\odot}$, albeit with
over an order of magnitude uncertainty.  For reference, 
\citet{2018ApJ...858...20V} reported a
3$\sigma$ upper limit on any potential black hole in this UCD galaxy to be less than 
$1.0\times10^5~M_{\odot}$, suggesting that they may have just missed out on
the required spatial resolution to detect the suspected black hole in this
UCD galaxy. 

\citet[][, their Section~6.1]{2018ApJ...858...20V} wrote that UCD~330, 
with its $M_{\rm bh}/M_{\rm UCD}$ mass ratio of just a few percent, 
is clearly different from other UCD galaxies with $M_{\rm bh}$--to--$M_{\rm
  UCD}$ mass ratios with percentages in the teens.  However, it makes more
sense to compare the $M_{\rm bh}$--to--$M_{\rm inner-cpt}$ mass ratio when
testing the galaxy threshing scenario.  Given that the 
$M_{\rm bh}/M_{\rm nc}$ ratio is not constant across nuclear clusters of
different masses, different  $M_{\rm bh}/M_{\rm inner-cpt}$ mass ratios in
UCD galaxies --- if they are stripped nuclei --- is expected.
These ratios are reported in Table~\ref{Tab1}. 

For UCD~320, \citet{2018ApJ...858...20V} reported a single-component structure (perhaps indicative of
a globular cluster rather than a UCD galaxy), with $M_V = -10.39$~mag and $M_*/L_V =
2.64$, and therefore one obtains a total stellar mass of 
$3.17\times10^6~M_{\odot}$.  If this is a remnant nuclear star cluster, then
from equation~\ref{Eq_new-sigma} 
one may expect it to contain a black hole with a mass equal to
$5.5\times10^4~M_{\odot}$, i.e.\ 20 times less than the 3$\sigma$ upper limit
of $10^6~M_{\odot}$ reported in \citet{2018ApJ...858...20V}. 

\begin{figure}
 \includegraphics[angle=-90, trim=2.2cm 2cm 1cm 9cm, width=\columnwidth]{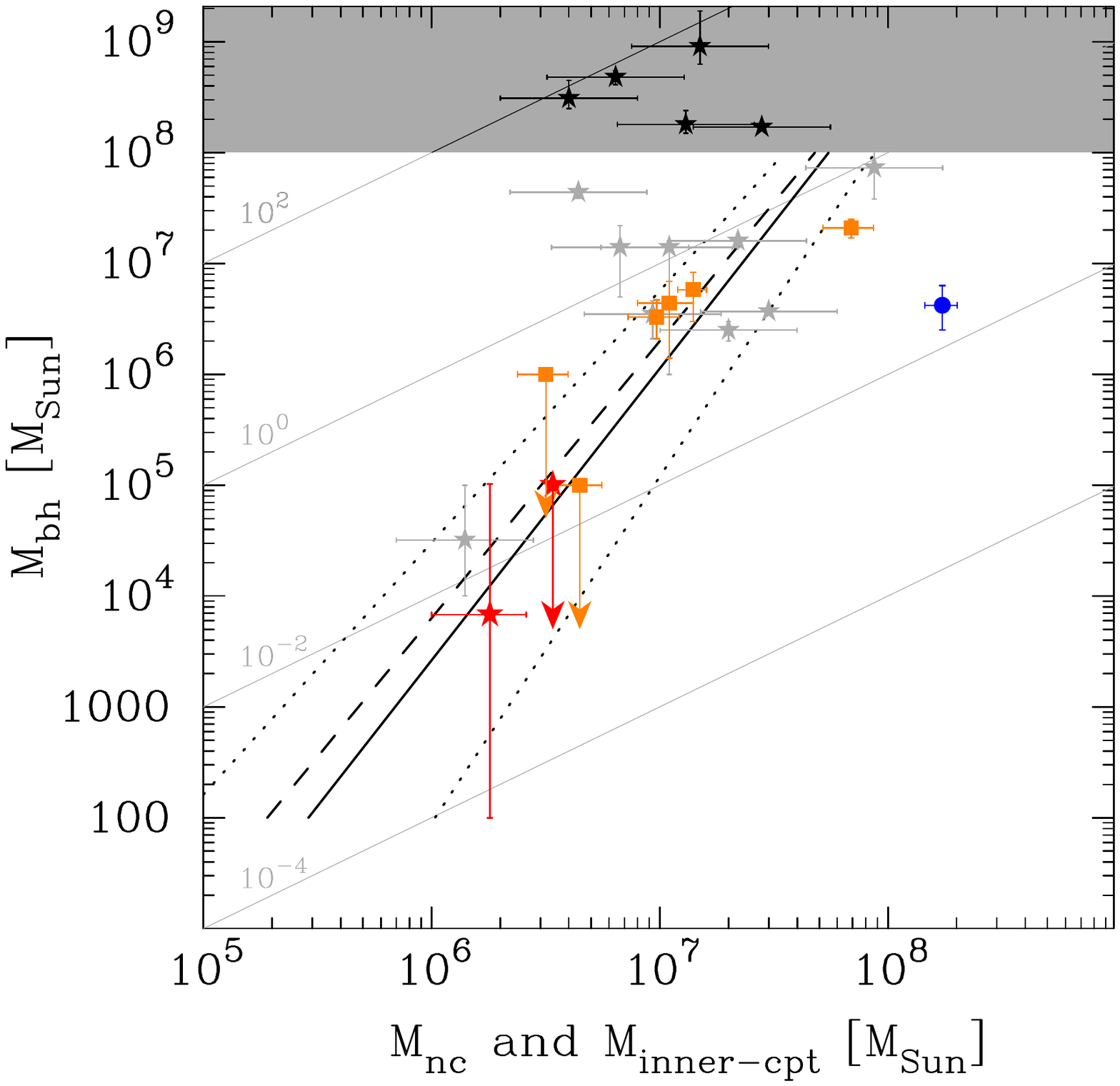}
    \caption{$M_{\rm bh}$--$M_{\rm nc}$ relations:  
      equation~\ref{Eq_mass} (dashed line) and 
    equation~\ref{Eq_sigma} (solid line, with the 1-sigma uncertainty 
    denoted by the dotted lines).  The data points represent: 
the inner (stellar mass) component of six UCD galaxies, thought to be remnant
nuclear star clusters (orange squares); 
the inner (stellar mass) component of one UCD galaxy (M59-UCD3), which may be
a remnant nuclear disc (blue circle); and the 
nuclear star cluster in NGC~205 and NGC~404 (2 red stars).  Three of these entries only
have upper limits on their black hole mass. 
A $\pm$25 percent uncertainty has been assigned to those stellar 
mass measurements without an error bar in Table~\ref{Tab1}.
The 14 small black and grey stars are nuclear star clusters with dynamically
measured black hole masses \citep{2009MNRAS.397.2148G, 2012MNRAS.422.1586G}. 
In galaxies with black hole masses greater than $\approx$$10^8\,{\rm M}_{\odot}$, the nuclear
  star clusters start to erode at the expense of the black holes 
\citep{2010ApJ...714L.313B}. 
}
\label{Fig2}
\end{figure}

For convenience, the above mentioned masses for all of the UCD galaxies are provided in
Table~\ref{Tab1} and shown in Figure~\ref{Fig2}.  One can immediately
appreciate that the measured  (black hole)-to-(inner star cluster) mass ratios
 for some of the UCD galaxies would make it challenging to separate the contributions
from the black hole and star cluster to the inner root-mean-square velocity
$V_{\rm rms}=\sqrt{sigma^2 + V_{\rm rot}^2}$ spike. 
The black hole's  
gravitational sphere-of-influence would loom large when the cluster's stellar
mass is just 2 to 3 times greater than the black hole's mass.  
The structure and motion of these inner star clusters will be partly dictated by the black hole
\citep{1976ApJ...209..214B, 1977ApJ...216..883B, 2015ApJ...814...57M}. 

%
%

\begin{table*}
\centering
\caption{Black hole and star cluster masses.}
\begin{tabular}{llllll}
\hline\hline
Object      & $M_{\rm bh,obs}$               & $M_{\rm nc,obs}$                & $M_{\rm bh,obs}$/$M_{\rm nc,obs}$  & $M_{\rm nc,pred}$                 &  $M_{\rm bh,pred}$  \\
            &    $M_{\odot}$                &     $M_{\odot}$                &                               &    $M_{\odot}$                  &    $M_{\odot}$  \\
 (1)  &  (2)  &  (3)  &  (4)  &  (5)  &  (6)  \\
\hline  
M60-UCD1    & $(2.1\pm0.4)\times10^7$      &  $6.9\times10^7$              &         0.30                 &  $(3.1^{+7.6}_{-2.2})\times10^7$  & ... \\
Virgo-UCD3  & $(4.4^{+2.5}_{-3.0})\times10^6$ &  $(1.1\pm0.3)\times10^7$      &         0.40                 &  $(1.7^{+4.4}_{-1.2})\times10^7$  & ... \\
M59cO (UCD) & $(5.8^{+2.5}_{-2.8})\times10^6$ &  $(1.4^{+0.2}_{-0.2})\times10^7$ &         0.41                 &  $(1.9^{+4.7}_{-1.4})\times10^7$  & ... \\
M59-UCD3$^a$ & $(4.2^{+2.1}_{-1.7})\times10^6$ &  $(1.73\pm0.28)\times10^8$   &         0.02$^a$              &            ...                 & ... \\
Fornax-UCD3 & $(3.3^{+1.4}_{-1.2})\times10^6$ &  $9.7\times10^6$              &         0.34                  &  $(1.5^{+3.9}_{-1.1})\times10^7$  & ...   \\
CenA-UCD330 & $<1.0\times10^5$              &  $4.46\times10^6$            &        $<0.02$                &           ...                  &  $1.3\times10^5$ \\
CenA-UCD320 & $<1.0\times10^6$              &  $3.17\times10^6$            &        $<0.32$                &           ...                  &  $5.5\times10^4$  \\
NGC~205     & $(6.8_{-6.7}^{+95.6})\times10^3$ &  $(1.8\pm0.8)\times10^6$     &        0.004                  &           ...                  &  $1.2\times10^4$  \\
NGC~404     & $<1.5\times10^5$              & $(3.4\pm0.2)\times10^6$      &        $<0.05$                &           ...                  &  $6.3\times10^4$  \\
NGC~3319    &     ...                       & $\approx$$6\times10^6$       &         ...                   &           ...                  &  $\approx$$3\times10^5$ \\
\hline
\end{tabular}

\label{Tab1}
Column 2: Observed, measured black hole mass.  
Column 3: Observed, measured star cluster mass; either the inner-component of
UCD galaxies or the nuclear star clusters of (non-stripped) galaxies. 
References to these masses are provided in the relevant subsections. 
Column 5: Predicted nuclear cluster mass based on $M_{\rm bh,obs}$ and equation~\ref{Eq_sigma}. 
Column 6: Predicted black hole mass based on $M_{\rm nc,obs}$ and equation~\ref{Eq_new-sigma}. 
$^a$ M59-UCD3 may represent a threshed nuclear disc.  The uncertainty on the
predicted black hole masses are large: $2.62\times0.5$ dex. 
\end{table*}

\section{Nuclear Star Clusters}\label{Sec_nc}

\subsection{NGC 205}

Having provided the above analysis for all five UCD galaxies with a directly
measured black hole mass to date, things shall now be reversed. That is,
rather than investigating if a reported black hole mass is consistent or not
with the speculation that it resides in a remnant nuclear star cluster, the
scaling relations will now be used to predict the mass of a black hole within
two bona fide nuclear star clusters which have just had their central black
hole masses measured.  However, because the shallow slope ($0.38\pm0.06$) of
the $M_{\rm nc}$--$M_{\rm bh}$ relation (equation~\ref{Eq_sigma}) corresponds
to a steep slope ($\approx 2.62\pm0.42$) for the $M_{\rm bh}$--$M_{\rm nc}$
relation, the uncertainty on $M_{\rm bh}$ is notably greater for a given
measurement of $M_{\rm nc}$.  As such, the predictive power with this approach
is less certain.  Despite this, because this method of scientific analysis is
so readily employable, yet has not been widely explored, a first Case Study
shall be made using the Local Group galaxy NGC~205.  In essence, this will
test if the black hole and nuclear star cluster in NGC~205 follow the $M_{\rm
  nc}$--$M_{\rm bh}$ relation used here.

The NGC~205 dwarf early-type galaxy has a well-resolved nuclear star cluster
with a stellar mass of $(1.4\pm0.7)\times10^6~M_{\odot}$
\citep{2006MNRAS.369.1321D, 2009MNRAS.397.2148G}, or more recently
$(1.8\pm0.8)\times10^6~M_{\odot}$ \citep{2018ApJ...858..118N}.

Bypassing any need to know the velocity dispersion of NGC~205's main
spheroidal component, which is distinct from the velocity dispersion of
NGC~205's nuclear star cluster, equation~\ref{Eq_sigma} can be inverted to
estimate the expected black hole mass in NGC~205's star cluster. This equation
for the black hole mass simply depends on the star cluster mass, and is such that
\begin{equation}
\log(M_{\rm bh}/M_{\odot}) = (2.62\pm0.42)
\log(M_{\rm nc}/[10^{7.83}\,M_{\odot}]) + (8.22\pm0.20).
\label{Eq_new-sigma}
\end{equation} 

The above equation gives an expected black hole mass in NGC~205 equal to
$6.4\times10^3~M_{\odot}$ if $M_{\rm nc} = 1.4\times10^6~M_{\odot}$, or
$M_{\rm bh}=1.2\times10^4~M_{\odot}$ if $M_{\rm nc} =
1.8\times10^6~M_{\odot}$.  This expectation from equations established a few
years ago is in remarkable agreement with \citet{2019ApJ...872..104N} who
report a dynamically-determined black hole mass in NGC~205 of
$6.8_{-6.7}^{+95.6}\times10^3~M_{\odot}$ ($\pm$3$\sigma$ uncertainty),
consistent with the upper limit of $3.8\times10^3~M_{\odot}$ ($\pm$3$\sigma$
uncertainty) from \citet{2005ApJ...628..137V}.

The agreement may reveal that the 
$M_{\rm nc}$--$M_{\rm bh}$ relation can be extended to black hole masses less
than $10^4~M_{\odot}$, or it may be fortuitous. 
While the expected black hole mass is more than two orders of magnitude less
than the nuclear star cluster mass, the uncertainty on the expected black hole
mass is approaching a factor of 40 --- dominated by the assigned intrinsic scatter (1.31
dex) which is
2.62 times greater along the $\log M_{\rm bh}$ direction than compared with the
$\log M_{\rm nc}$ direction.

\subsection{NGC~404}

One can similarly ask the question: What black hole mass is expected in the
nuclear star cluster of the nearby dwarf galaxy NGC~404
\citep{2010ApJ...714..713S}.  \citet{2017ApJ...836..237N} constrain the mass
of this black hole to be less than $1.5\times 10^5~M_{\odot}$.  From the
stellar mass $(3.4\pm0.2)\times10^6~M_{\odot}$ of this galaxy's innermost
component, referred to as the ``Central excess S\'ersic'' by \citet[][, see
  their Table~4 and Figure~10]{2017ApJ...836..237N}, one can predict the
expected black hole mass.  This approach differs from \citet[][, their
  Section~6.2]{2017ApJ...836..237N} who refer to the inner two components as
the nuclear star cluster.  From equation~\ref{Eq_new-sigma}, one expects
$\log(M_{\rm bh}/M_{\odot})=4.8\pm1.5$ ($M_{\rm bh}=6.3\times10^4~M_{\odot}$)
if the intrinsic scatter is 2.62$\times$0.5~dex.  This total uncertainty of
1.5~dex, which was calculated in quadrature from the various smaller
uncertainties, would be halved if the intrinsic scatter was excluded.

This is an intriguing case study as the ``Central excess S\'ersic'' is thought to be
1~Gyr old and accreted \citep{2017ApJ...836..237N}, 
while the second (larger) component has a 
half-light radius of 20~pc 
and would be considered a nuclear disc by \citet{2007ApJ...665.1084B} and 
\citet{2013ApJ...763...76S} due
to its size.  As such, the $M_{\rm nc}$--$M_{\rm bh}$ equations used here are
not applicable to the second component.  
In passing, it is noted that UCD1 around NGC~4546 was also forming stars just
1 to 2 Gyr ago \citep{2015MNRAS.451.3615N}. 
The topic of the stellar populations in nuclear star clusters
\citep[e.g.][]{2001AJ....121.1473B, 2009AN....330..969P, 2016MNRAS.463.1605L},  
is, however, beyond the scope of this paper.

\subsection{Other} 

As for additional nuclear star clusters that have been suggested to house an
intermediate mass black hole, \citet{2018ApJ...869...49J} 
report an X-ray point-source in the barred spiral galaxy
NGC 3319. They find that the X-ray SED is more consistent with an AGN rather than a
super-Eddington, stellar-mass, ultra-luminous X-ray source.  They find the
X-ray point-source to be coincident with 
the nuclear star cluster, whose mass they report to be around
$6\times10^6~M_{\odot}$. From equation~\ref{Eq_new-sigma}, the expected black
hole mass is $3\times10^5~M_{\odot}$.  

The Virgo cluster, dwarf early-type galaxies, IC~3442 and IC~3292 similarly
have X-ray point-sources coincident with nuclear star clusters that are
expected to house $\sim$$10^5~M_{\odot}$ 
black holes
\citep{2019MNRAS.484..794G}.  Pushing to yet lower masses, the early-type
galaxies IC~3602 and IC~3633 have been predicted to house black hole masses of
$\sim$$10^4~M_{\odot}$  
\citep{2019MNRAS.484..794G}, although this is based on
their galaxy luminosity and velocity dispersion rather than a nuclear cluster
mass or X-ray emission.

The spiral galaxy NGC~4178 may harbour an intermediate mass black hole
\citep{2009ApJ...704..439S, 2012ApJ...753...38S}, and an estimate of just
$10^3~M_{\odot}$ has been made \citep{2019MNRAS.484..814G}. 
However, this galaxy's
centrally-located, X-ray point-source has a soft, probably thermal, spectrum
suggestive of a stellar-mass black hole in a high/soft state.  Further
investigation is required.


\section{Discussion}\label{Sec_discuss}

There is a slight twist to all of this. 
One may ask, were the UCD galaxies once in early- or late-type galaxies, or in galaxies
with or without a substantial stellar disc?  
Some of the analysis performed here has implicitly assumed that the UCD galaxies were once
in galaxies with (intermediate- or large-scale) stellar discs. 
This is because early- and late-type galaxies
appear to follow 
different $M_{\rm bh}$--$M_{\rm *,bulge}$ and $M_{\rm
  bh}$--$M_{\rm *,galaxy}$ relations 
\citep{2016ApJ...817...21S, 2016ApJ...831..134V, 2017ApJ...844..170T,
  2019ApJ...873...85D, 2019ApJ...876..155S}. 
The use of equation~\ref{eq_Sersic} --- which is 
consistent with the $M_{\rm bh}$--$M_{\rm *,bulge}$ relation for both late-type
galaxies and early-type galaxies with either an intermediate- or a large-scale
disc \citep{2019ApJ...873...85D, 2019ApJ...887...10S} --- effectively creates a link to
galaxies with a substantial stellar disc (see Figure~\ref{Fig1}).  
One might consider use of the latest $M_{\rm bh}$--$M_{\rm *,bulge}$ relations
from \citet{2019ApJ...876..155S}, however this does not seem
particularly appetising given the apparent order of magnitude difference in black hole
mass depending on whether the original early-type galaxies 
contained a disc or not.  Given that the $M_{\rm bh}$--$M_{\rm
  *,bulge}$ relations depend on galaxy type, future progress would benefit
from establishing if the $M_{\rm nc}$--$M_{\rm *,bulge}$ relations also depend
on either the galaxy type (see Figure~5 in \citet{2016MNRAS.457.2122G} 
in regard to the $M_{\rm nc}$--$M_{\rm *,gal}$ relations)  
or the presence/absence of a substantial disc.  Establishing and 
partnering these (morphological type)-specific 
relations should shed further light on this topic and yield improved
predictive power.  
However, our preferred use of the scaling relations involving the
velocity dispersion may have circumvented this issue. This is because 
 nucleated\footnote{\citet{2019ApJ...887...10S} find that
  (non-nucleated) core-S\'ersic early-type galaxies and (typically-nucleated)
  S\'ersic early-type galaxies define different $M_{\rm bh}$--$\sigma$
  relations.}  early- and late-type S\'ersic galaxies may follow the same
$M_{\rm bh}$--$\sigma$ relation \citep{2019ApJ...887...10S}.  
Whether they follow different $M_{\rm nc}$--$\sigma$ relations is yet to be
established. 

Another implicit assumption 
has been that the local ($z\approx0$) relation between nuclear star 
cluster mass and black hole mass held when the UCD galaxy was formed.  This
may not have been so. 
However, given the good agreement between the UCD galaxies and the local 
(black hole)--(nuclear star cluster) mass scaling relation, there may have been little evolution in this 
relation, which does not discount evolution along the relation for those nuclear
star clusters and black holes that remained at the centre of a galaxy. 
The distribution in Figure~\ref{Fig2}, showing the mass of the UCD galaxies' black
hole and inner stellar component, supports the notion that UCD galaxies are
the remnant nuclei of threshed galaxies.


The increased number density of black holes at the low-mass
end of the black hole mass function \citep[e.g.][]{2004MNRAS.354.1020S,
  2007MNRAS.378..198G, 2009MNRAS.400.1451V, 2013ApJ...764...45K}, due to the (expected) 
central massive black holes in UCD galaxies \citep{2014MNRAS.444.3670P,
  2016MNRAS.458.2492P}, has a couple of immediate
consequences.  It will increase the expected number of extreme 
mass-ratio inspiral (EMRI) events --- involving stellar mass black
holes and neutron stars merging with the massive black hole ---  and therefore
increase the expected number of gravitational wave events 
\citep[e.g.][]{2007CQGra..24R.113A, 2010PhRvD..81j4014G, 2010ApJ...718..739M,
  2018ApJ...867..119F}.  As detailed in \citet{2012A&A...542A.102M}, 
the steeper (super-quadratic rather than near-linear) 
$M_{\rm bh}$--$M_{\rm bulge}$ scaling relation implied an order of magnitude
reduction to the number of 
EMRI events that the Laser Interferometer Space Antenna 
\citep[LISA:][]{2017arXiv170200786A, 2019arXiv190706482B} could detect 
in galactic nuclei hosting nuclear star 
clusters and massive black holes.  As such, the promise of additional EMRI events from
UCD galaxies is likely to be 
exciting and welcome news for the new LISA mission and Taiji program 
\citep{2018arXiv180709495R}. 

Second, it will enhance 
the expected number of stellar tidal disruption events 
\citep[TDE:][]{1975Natur.254..295H, 2013IAUS..290...53K, 2016MNRAS.455..859S} 
given the abundance of UCD galaxies 
\citep{2006AJ....131..312J, 2010ApJ...722.1707M, 2011ApJ...737...86C,
  2019ApJ...871..159V}.  Third, and related, 
it opens the question as to whether UCD galaxies may simmer away due to weak,
sporadic, AGN activity.  The massive black hole in the Milky Way's nuclear
star cluster\footnote{Excitingly, the Milky Way's nuclear star cluster seems
  set to be monitored in the near-infrared by the planned Japanese satellite 
  Small-JASMINE \citep{2013IAUS..289..433Y, 2018IAUS..330..360Y}.}  
sputters and bubbles according to its time-variable fuel supply.  This
is evidenced by both the daily/yearly variations in flux at different
wavelengths  \citep[e.g.][]{1981ApJ...248L..13R, 2001ApJ...547L..29Z,
  2001Natur.413...45B, 2003Natur.425..934G, 2004ApJ...601L.159G,
  2005ApJ...623L..25M} 
and the $\sim$200~pc radio bubble discovered by \citet{1984Natur.310..568S}, 
see also \citep{1985PASJ...37..359T}, 
with known extensions to higher Galactic latitudes
\citep{1977A&A....60..327S, 2000ApJ...540..224S, 2003ApJ...582..246B,
  2010ApJ...724.1044S, 2019Natur.573..235H}. 
The presence of weakly-active massive black holes in UCD galaxies may further diminish
prospects for life in such systems \citep[e.g.][]{2001Icar..152..185G,
  2013AsBio..13..491J, 2013MNRAS.431...63T, 2018ApJ...864..115K}, but see
\citet{2016ApJ...827...54D}. 
This is especially true given the close proximity 
of the stars and potential planets to the massive black hole 
\citep{2018ApJ...855L...1C, 2019arXiv191000940S}.



\section*{Acknowledgements}

The result for M60-UCD1  was first presented by the author at the Wilhelm and Else Heraeus Seminar
631 titled ``Stellar aggregates over mass and spatial scales'', and held over
5-9 Dec.\ 2016.  
The author is grateful for their support and pleased to finally publish that result
here. 
This research was additionally supported under the Australian Research Council's funding
scheme (DP17012923).



\bibliographystyle{mnras}
\bibliography{Graham-UCD1}{}

\bsp    
\label{lastpage}
\end{document}